\documentclass[preprint,authoryear,11pt]{elsarticle}
\usepackage[margin={2.54cm,2.54cm}]{geometry}
\usepackage{float}

\usepackage{graphics}
\usepackage{graphicx}
\usepackage{epsfig}
\usepackage{color}
\usepackage{amsmath}
\usepackage{amssymb}
\usepackage{pifont}
\input{tcilatex}

\def\div{{\rm div}}

\journal{Journal of the Mechanics and Physics of Solids}

\begin{document}

\begin{frontmatter}



\title{A comparative analysis of numerical approaches to the mechanics of elastic sheets}


\author[harvard]{Michael Taylor}
\author[UMass]{Benny Davidovitch}
\author[UMass]{Zhanlong Qiu}
\author[harvard,kavli]{Katia Bertoldi}
\address[harvard]{School of Engineering and Applied Sciences, Harvard University, Cambridge, MA}
\address[UMass]{Physics Department, University of Massachusettes Amherst}
\address[kavli]{Kavli Institute for Bionano Science and Technology, Harvard University,
Cambridge, MA}

\begin{abstract}
Numerically simulating deformations in thin elastic sheets is a challenging problem in computational mechanics due to destabilizing compressive stresses that result in wrinkling. Determining the location, structure, and evolution of wrinkles in these problems have important implications in design and is an area of increasing interest in the fields of physics and engineering. In this work, several numerical approaches previously proposed to model equilibrium deformations in thin elastic sheets are compared. These include standard finite element-based static post-buckling approaches as well as a recently proposed method based on dynamic relaxation, which are applied to the problem of an annular sheet with opposed tractions where wrinkling is a key feature. Numerical solutions are compared to analytic predictions, enabling a quantitative evaluation of the predictive power of the various methods. Results indicate that static finite element approaches are highly sensitive to initial imperfections, relying on \textit{a priori} knowledge of the equilibrium wrinkling pattern to generate optimal results. In contrast, dynamic relaxation is much less sensitive to initial imperfections and can generate solutions for a wide variety of loading conditions without requiring knowledge of the equilibrium solution beforehand.
\end{abstract}


\end{frontmatter}


\section{Introduction \label{sec:Introduction}}

Thin elastic sheets are not only found abundantly in nature (\citealt{BenAmar2008}), but are also used in a wide variety of structural applications (\citealt{Carbonez}) because of their excellent tensional resistance-to-weight ratio. Determining equilibrium deformations in these structures is nontrivial as loading a thin sheet typically results in regions that are locally tense, compressed, or slack (i.e., stress-free). In the compressed regions, wrinkles form in response to instability.  Wrinkles may be something engineers wish to avoid (e.g., in solar sails, \citealt{Vulpetti2008}) or perhaps something that can be used to control membrane behavior (\citealt{Vandeparre2010, Breid2013}). In either case, it is of great interest to be able to determine their structure (i.e. amplitude and wavelength) and location in a sheet at a given applied loading state.

Theoretical investigations of deformation and tensional wrinkling in thin sheets go back to the works of \citet{Wagner1929} and \citet{Reissner1938}.  Early research focused on assuming the sheet to be perfectly flexible and using membrane theory to model its deformation (\citealt{Mansfield1968,SteinHedge1961}). These studies formed the foundation of tension-field theory (\citealt{Pipkin1986,Steig1990,HaseSteig1994}), in which the stress field at the midplane of the sheet is assumed to have no compressive components. Tension-field theory is the appropriate leading-order model at vanishing thickness (\citealt{Pipkin1986,LDRaoult1995}) and is much more tractable than shell (or higher order) models incorporating bending stiffness (which involve computation of the curvature and its derivatives). However, while tension-field-theory is very useful for assessing the stress distribution and {\emph{location}} of wrinkled regions in very thin sheets (\citealt{HaseSteig1994,Taylor2014}), it offers no information on the actual  {\emph{structure}} of wrinkles. Here, we call the membrane-dominant regime where tension-field theory is valid the ``far-from-threshold (FT)" parameter regime.  In this regime, the compression induced by the tensile loads is much larger than the thickness-dependent level at which a real sheet buckles. Thus, tension-field-theory cannot describe the {\emph{evolution}} of a real sheet, upon increasing loads, from a bending-dominant (or "near-threshold" (NT)) regime at the onset of wrinkling to the FT regime, at which a fully wrinkled pattern develops.

In the last decade, several groups have attempted to develop a comprehensive framework that addresses simultaneously the {\emph{location, structure}} and {\emph{evolution}} of tensional wrinkle patterns, focusing on a few basic set-ups, such as a rectangular sheet under stretch (\citealt{Friedl2000,Nayyar2011,Puntel2011,CRM2002,CM2003,Nayyar2011,Healey2013}) or shear (\citealt{WP2006a}, \citeyear{WP2006b}, \citeyear{WP2006c}; \citealt{Zheng2009,Diaby2006}), a disk-like (\citealt{King2012}) or annulus-like sheet (\citealt{Bella2000,Geminard2004,Coman2007,Coman2007b,Davidovitch2011,Davidovitch2012,Pineirua2013,Toga2013}) under axially-symmetric tensile loads, and a stretched-twisted ribbon (\citealt{Chopin2013,Chopin2014}). In particular, the first of these examples has attracted considerable interest in the mechanical engineering community (\citealt{Nayyar2011}). Here, a rectangular-shaped sheet is stretched where its short edges (of width $W$) are clamped and its long edges (of length $L$) are free to contract, such that a pattern of parallel wrinkles emerges in a large portion of the sheet, away from the clamped edges. An early numerical work (\citealt{Friedl2000}) has focused on the onset of wrinkles ({\emph{i.e.}} the NT regime) in this system, and results were interpreted by drawing analogy to the classical Euler buckling of rods. Later, \cite{CM2003} addressed the {\emph{structure}} of this tensional wrinkling pattern away from threshold ({\emph{i.e.}} in the FT regime) by drawing an analogy to the elementary example of uniaxially compressing a rectangular sheet on a substrate of stiffness $K$, which is known to exhibit parallel wrinkles of wavelength $\lambda \approx (B/K)^{1/4}$, where $B$ is the bending modulus of the sheet and $K$ is the substrate's stiffness. The essential observation (\citealt{CRM2002,CM2003}) was that the presence of tension $T$ along wrinkles of length $L$ induces an effective  substrate of stiffness $K=T/L^2$, such that the wavelength of tensional wrinkles satisfies the scaling law: $\lambda \sim (B/T)^{1/4} L^{1/2}$.

Subsequent works by several groups, which addressed the stretched rectangular sheet, have attempted to describe the complete evolution of the wrinkle pattern, as the tensile load is gradually increased from its threshold value to the FT behavior addressed in (\citealt{CM2003}). These studies, however, were encountered by significant difficulties: experimental efforts to probe the onset of the wrinkling instability (\citealt{Zheng2009}) were baffled by the high sensitivity of this system to the non-uniformity of the applied loads and the likelihood of plastic deformations on various scales; on the theoretical front -- the difficulty may be attributed to the lack of analytic solutions of the stress field neither for the planar state (necessary to describe the onset of instability and the NT regime), nor for tension field theory (which provides the basis for analysis of the FT regime). This situation highlights the important role of numerical simulations, even in such a basic system, as the ultimate route for a systematic study of tensional wrinkling. The computational challenge here stems from the multi-scale nature of wrinkling phenomena, whereby the wavelength $\lambda$ vanishes with the sheet's thickness, while the size of the wrinkled region is determined by the length $L$ of the sheet.

\begin{figure}[h]
\begin{centering}
\includegraphics[scale=0.6]{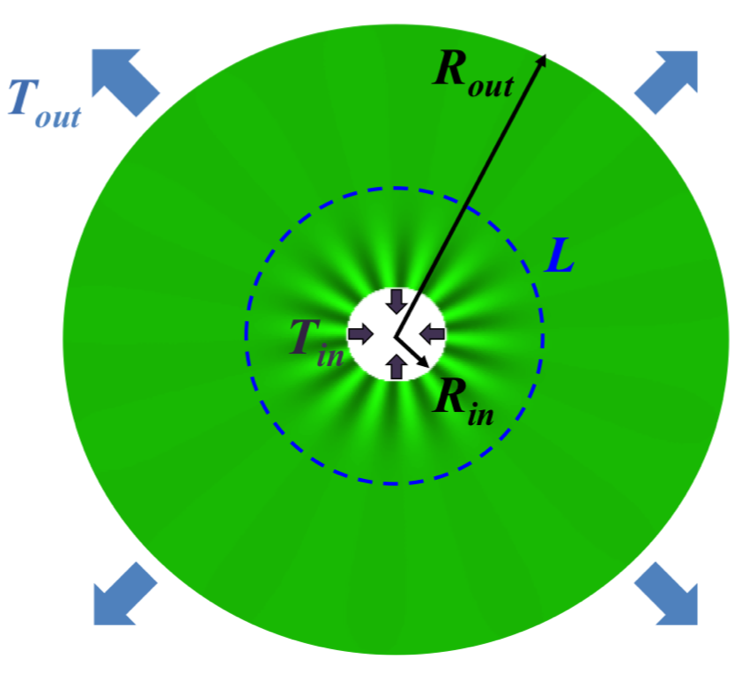}
\par\end{centering}

\caption{Classical Lam\'e set-up. Opposing tractions $T_{in}$ and $T_{out}$ applied to the inner an outer boundary, respectively, of the annulus causes compressive stresses in the region $R_{in}<r<L$ leading to wrinkling.}
\label{fig:Setup}
\end{figure}

Recognizing the need in reliable, efficient numerical simulations, the primary purpose of this paper is to examine and quantitatively compare the performance of some popular numerical methods for studying the key aspects of tensional wrinkling patterns -- their {\emph{location, structure}}, and {\emph{evolution}} as the tensile loads are being varied. This purpose dictates our choice of case study, which is known as the Lam\'e problem (\citealt{Timoshenko1970}): an annular sheet under radial tensile loads $T_{in}$ and $T_{out}$, exerted, respectively, on its inner and outer boundaries (see Figure \ref{fig:Setup}). The key advantage of the Lam\'e set-up, in comparison to a stretched rectangular sheet, is the existence of analytic predictions for the location and structure of the wrinkle pattern in both NT and FT regimes, as well as the evolution of the pattern between these parameter regimes as the tensile loads are gradually increased. A solution to the stress distribution of the planar state, which can be found in classical textbooks on elasticity theory  (\citealt{Timoshenko1970}) has been attributed to Lam\'e, and has been recently used for linear stability analysis that yields the threshold value of the tensile loads (\citealt{Coman2007b}), as well as the location and structure of the wrinkle pattern in the NT regime. More recently, an exact solution of the tension-field theory equations has been obtained (\citealt{Coman2007,Davidovitch2011}), allowing one to identify exactly the location of the wrinkled zone in a very thin sheet, away from threshold.

The structure of the wrinkle pattern in the corresponding FT regime was described through a singular expansion of the F\"oppl-von K\'arm\'an (FvK) equations around the tension-field-theory stress field, and the evolution from the NT and FT regimes was characterized (\citealt{Davidovitch2012}). This progress provides us with nontrivial analytic results on the location, structure, and evolution of the wrinkle pattern upon varying the tensile loads, which can be used to test the reliability of simulation methods.

Focusing on the Lam\'e set-up, and comparing numerical results of three distinct simulation schemes with the analytic predictions, our main result is summarized in Table \ref{table:Comparison}. The first two approaches use the finite element method with general purpose shell elements as implemented in Abaqus/Standard to determine the buckling eigenmodes, where the first method uses the buckling modes as an imperfection in a general static analysis, and the second method uses the buckling modes as an imperfection in a Riks analysis. The third approach involves the dynamic relaxation method that has been recently used to solve for equilibrium deformations in thin sheets exhibiting wrinkling (\citealt{Taylor2014}).  All three methods are based (at least in part) upon Koiter's shell model, differing primarily in how the equations are solved.
As Table \ref{table:Comparison} shows, both finite element approaches can accurately predict the stress field (and hence the {\emph{location}} of wrinkles), but are unable to determine the correct wrinkle state ({\emph{i.e.}} structure) unless the analyst has \textit{a priori} knowledge of the appropriate wrinkling modes.  In contrast to the first two methods, the dynamic relaxation approach is able to accurately predict both the location and structure of the wrinkle morphology, as well as the evolution of the stress field and the pattern from the planar state, through the NT paramer regime, to the fully developed, FT state of the system.

\begin{table}[h]
\begin{center}
\begin{tabular}{l | *{2}{c}}

Method                    & Macro-scale Properties & Micro-scale Properties \\
\hline
Finite Element: Static    & \checkmark & \ding{53} \\
Finite Element: Riks      & \checkmark & \ding{53} \\
Dynamic Relaxation        & \checkmark & \checkmark \\
\end{tabular}
\end{center}
\caption{Comparison of solution methods and general capabilities in the FT regime for the considered case study}
\label{table:Comparison}
\end{table}

The structure of the paper is as follows. In Section 2, we motivate the use of Koiter's model for tensional wrinkling problems and provide a brief summary of the relevant equations. In Section 3, we provide an overview of the three numerical approaches used in this study.  Finally, in Section 4, we describe the Lam\'e problem, provide numerical results, and compare them.

\section{Theoretical description of tensional wrinkling}
Before turning to a detailed description of the numerical methods and our simulation results, let us make some general comments on the theoretical description of tensional wrinkling phenomena. The energy of a homogenous, isotropic elastic sheets, can be described as a sum of a ``membrane" term and a ``bending" term, which depend, respectively, on the strain and curvature of the mid-plane. For a sheet of thickness $t$ made of elastic material of Young modulus $E$, these two energies are proportional, respectively, to the stretching modulus ($Y \sim Et $) and bending modulus ($B \sim Et^3$). At a fundamental level, the success of tension-field theory in describing the location of wrinkles in very thin sheets stems from the dominance of the membrane term, whereas its inability to predict structure stems from ignoring the higher order terms in the energy, which involve a bending cost. Thus, a consistent theoretical framework of tensional wrinkling phenomena must account for both membrane and bending terms. The FvK theory is a particularly popular model that accounts for both membrane and bending terms, and has been used successfully to describe wrinkling phenomena in elastic sheets (\citealt{Davidovitch2011, Davidovitch2012, Chopin2014}). Specifically, the theoretical study of the Lam\'e problem (\citealt{Davidovitch2012}), whose predictions we use in our work, is essentially a non-perturbative analysis of FvK equations that addresses both NT and FT parameter regimes. Nevertheless, from the perspective of elasticity theory of solid bodies in 3D, the FvK model is  merely a successful phenomenological model, since certain  assumption on the stress distribution across the sheet's thickness must be made. This leads one to the question of what is the best theory to use for modeling deformation in thin sheets where wrinkling is a key feature of interest.

In recent years, research in plate and shell theory has focused on placing existing two-dimensional theories onto rigorous mathematical foundations (\citealt{Friesecke2006,Ciarlet2000}) as well as generating new theories of optimal accuracy with respect to the parent three-dimensional elasticity theory (\citealt{Friesecke2006,Steig2013b}). Both asymptotic (\citealt{Ciarlet1980}) and gamma-convergence (\citealt{Friesecke2006}) methods have been used successfully to place, for instance, membrane, Kirchoff-Love, and the FvK theories on solid theoretical foundations given particular assumptions on the energy. For example, the FvK equations can be rigorously derived from three-dimensional elasticity theory only if the energy (per unit area) scales as $h^5$ and if the midplane deforms isometrically (\citealt{Friesecke2006}).  Notably, gamma-convergence and asymptotic methods have not yet yielded a model with both stretching and bending in a single framework (\citealt{Steig2008}).

However, Koiter's nonlinear shell theory (\citealt{Koiter1966}) is unique in that it wasn't developed as a limit model but rather based on considerable mechanical understanding and intuition (\citealt{Ciarlet2000}). \cite{Ciarlet2005} advocated this model as the best model in both the membrane-dominated and bending-dominated regimes in the case of linear elastic deformations. Hilgers and Pipkin (\citeyear{HPip1992,HPip1992b,HPip1996}) initiated the work to rigorously extend Koiter's model to finite mid-surface strains but used an ad hoc regularizing term to deal with destabalizing compressive stresses.  Steigmann (\citeyear{Steig2008,Steig2010,Steig2013b}) continued this work and derived a well-posed and optimally accurate (in terms of the parent three-dimensional theory) finite-strain shell model free of the ad hoc term.  In addition, \cite{Steig2013} showed that Koiter's model is the leading-order energy in the intermediate regime where bending and stretching deformation energies are assumed to be of comparable importance. This is precisely the region of interest in tensional wrinkling problems.

\subsection{Koiter's nonlinear isotropic plate model}
Here, we briefly summarize Koiter's model for an initially flat isotropic elastic plate (\citealt{Steig2013}). In the following, we note that Latin indices take values from \{1,2,3\} and Greek indices take values from \{1,2\}. The implementation of this model in both Abaqus and the dynamic relaxation codes is discussed in Section 2.2.

Stretch-induced wrinkling is associated with deformations where the stretching and wrinkling energies in the sheet are of comparable importance. In this regime, the leading order energy is (\citealt{Koiter1966,Steig2013,Taylor2014})
\begin{equation}
\label{eqn:eq2}
W=\tfrac{1}{2}h\{\tfrac{2\lambda \mu }{\lambda +2\mu }(tr\mathbf{\epsilon )}%
^{2}+2\mu \left\vert \mathbf{\epsilon }\right\vert ^{2}\}+\tfrac{1}{24}%
h^{3}\{\tfrac{2\lambda \mu }{\lambda +2\mu }(tr\mathbf{\kappa )}^{2}+2\mu
\left\vert \mathbf{\kappa }\right\vert ^{2}\},
\end{equation}
where $h$ is the sheet thickness, $\lambda$ and $\mu $ are the classical Lam\'{e} moduli, $\mathbf{\epsilon}=E_{\alpha\beta}\mathbf{e}_{\alpha}\otimes\mathbf{e}_{\beta}$ is the in-plane part of the Lagrange strain tensor and $\mathbf{\kappa}$ is the bending strain.  The bending strain is related to the deformed surface curvature tensor, $\mathbf{b}$ via
\begin{equation}
\mathbf{\kappa} = -(\nabla{\mathbf{r}})^T\mathbf{b}(\nabla{\mathbf{r}}),
\end{equation}
where $\textbf{r}$ is the position of a material point on the deformed image $\omega$, of the reference midplane, $\Omega$ and the superscript $T$ denotes tensor transpose. In components,
\begin{equation}
\mathbf{\kappa} = -b_{\alpha\beta}\mathbf{e}_{\alpha}\otimes\mathbf{e}_{\beta}; b_{\alpha\beta} = n_{i}r_{i,\alpha\beta},
\end{equation}
where $n_{i}$ are the components of the unit normal on the deformed surface, $ r_{i}$ are the components of the vector $\textbf{r}$, and the subscript comma refers to partial differentiation with respect to the coordinates.

The Euler equations associated with foregoing strain energy are
\begin{equation}
\label{eqn:eq3}
\div\mathbf{T}=\mathbf{0,}\text{\quad or\quad}T_{i\alpha,\alpha}=0.
\end{equation}
The tensor $\mathbf{T}$ is related to the three dimensional first Piola-Kirchoff stress evaluated on the mid-surface $\mathbf{P}$ via
\begin{equation}
\mathbf{T} = \mathbf{P}\mathbf{1},
\end{equation}
where $\mathbf{1} = \mathbf{I}-\mathbf{k}\otimes\mathbf{k}$, $\mathbf{I}$ is the three dimensional identity, and $\mathbf{k}$ is the unit normal to $\Omega$.  It has nontrivial components
\begin{equation}
\label{eqn:eq4}
T_{i\alpha}=N_{i\alpha}-M_{i\alpha\beta,\beta},
\end{equation}
where
\begin{equation}
\label{eqn:eq5}
N_{i\alpha}=\partial W/\partial r_{i,\alpha}\text{\quad and \quad}M_{i\alpha\beta}=\partial W/\partial r_{i,\alpha\beta},
\end{equation}
with $W$ given by (\ref{eqn:eq2}). Substituting (\ref{eqn:eq2}) into (7) yields,
\begin{equation}
M_{i\alpha\beta}=\tfrac{1}{12}h^{3}n_{i}\left(\tfrac{2\lambda\mu}{\lambda+2\mu}b_{\gamma\gamma}\delta_{\alpha\beta}+2\mu b_{\alpha\beta}\right).
\end{equation}
and
\begin{equation}
N_{i\alpha}=hr_{i,\beta}(\tfrac{2\lambda\mu}{\lambda+2\mu}E_{\gamma\gamma}\delta_{\beta\alpha}+2\mu E_{\beta\alpha}
)-M_{i\lambda\mu}\Gamma_{\alpha\lambda\mu},
\end{equation}
where $\Gamma_{\alpha\lambda\mu}$ are the Christoffel symbols. These are related to the gradients of the in-plane strain via,
\begin{equation}
\Gamma_{\alpha\lambda\mu}=E_{\mu\alpha,\lambda}+E_{\alpha\lambda,\mu}-E_{\lambda\mu,\alpha}.
\end{equation}

The edge boundary of the reference mid-plane $\Omega$ is denoted $\partial\Omega$. In general mixed-boundary problems, position and orientation data are assigned on a part of the boundary denoted $\partial\Omega_{e},$ and traction and bending moments assigned on a part of the boundary $\partial\Omega_{n}.$ Typical boundary conditions on $\partial\Omega_{e}$ entail the specification
of the position $\mathbf{r}$ and its normal derivative $\mathbf{r}_{,\nu}$. Typical boundary conditions on $\partial\Omega_{n}$ are (\citealt{Steig2010})
\begin{equation}
\label{eqn:eq6}
T_{i\alpha}\nu_{\alpha}-(M_{i\alpha\beta}\nu_{\alpha}\tau_{\beta}),_{s}=f_{i}\text{\quad and\quad}M_{i\alpha\beta}\nu_{\alpha}\nu_{\beta}=c_{i},
\end{equation}
where $f_{i}$ and $c_{i}$ are the force and couple per unit length, $\tau_{\beta}$ are components of the unit tangent to the boundary, and $s$ is the arc-length.  In this study, we consider boundaries where only $f_{i}$ and $c_{i}$ are specified, i.e. $\partial\Omega_{n}=\partial\Omega$.

We note that the extension of (\ref{eqn:eq2}) to finite midsurface strain is given by (\citealt{Steig2013})
\begin{equation}
\label{eqn:eq12}
W=\tfrac{1}{2}h\mathcal{W}+\tfrac{1}{24}%
h^{3}\{\tfrac{2\lambda \mu }{\lambda +2\mu }(tr\mathbf{\kappa )}^{2}+2\mu
\left\vert \mathbf{\kappa }\right\vert ^{2}\},
\end{equation}
where $\mathcal{W}$ is an appropriate strain energy function suitably restricted to plane stress. Problems with large midsurface strains are outside the scope of this work; however, they are common in polymeric and biological thin films (\citealt{CM2003}). Further, in contrast to the present theory, we note that the FvK theory assumes that all nonlinear terms of the Lagrange strain are negligibly small except those involving derivatives of the out-of-plane displacement.

\subsection{Numerical implementation of Koiter's model}
Koiter's model as described above is implemented directly in the dynamic relaxation code used in this study (more details can be found in Section 3.3).

For the numerical analysis in this work, we also use Abaqus/Standard 6.10 with S4R shell elements. The S4R element is a 4-node quadratic finite-membrane-strain element with reduced integration that has been used successfully in wrinkle analyses (\citealt{Zheng2009,Nayyar2011}) and is considered a robust, general purpose element (Abaqus 6.10 Theory Manual). The bending strains in these elements are computed based on the Budiansky-Sanders form (\citealt{Budiansky1962,Ciarlet2000}, Abaqus 6.10 Theory Manual) of Koiter's linear shell model (sometimes referred to as the Koiter-Sanders theory). The membrane strains are computed based upon a user-specified constitutive law. Although an explicit expression for the strain energy is not given in the Abaqus theory documentation, it is likely to be akin to (\ref{eqn:eq12}) but with the bending strain $\mathbf{\kappa}$ replaced with the linearized Budiansky-Sanders bending strain (\citealt{Budiansky1962,Ciarlet2000}). The status of the combined bending-stretching model used in the S4R element \textit{vis-\'a-vis} three dimensional elasticity theory is not known to the authors.

\section{Overview of numerical approaches \label{sec:Overview}}
In this section we describe the three numerical approaches that have been used to simulate equilibrium deformations in isotropic elastic sheets.  The first two are based on the finite element method as implemented in the commercial package, Abaqus. The third involves a specifically designed numerical code (\citealt{Taylor2014}), based on the method of dynamic relaxation.

The finite element method has been used in a variety of studies of wrinkling in thin sheets (\citealt{Zheng2009,WP2006c,Nayyar2011,Healey2013,Carbonez}). There are typically three approaches used: static post-buckling (\citealt{Zheng2009,WP2006c}), the Riks method (\citealt{Nayyar2011}), and explicit (\citealt{Zhang2013}). The first two rely on an initial analysis to determine the buckling eigenmodes of the sheet. The third approach, explicit dynamic simulation, is not attempted in this study as it very similar to the dynamic relaxation approach that is demonstrated here.

\subsection{Finite Elements: General Static Post-buckling \label{sec:FE_static}}
The first step in this procedure is to obtain the buckling eigenmodes and eigenfrequencies. The goal of this eigenvlaue buckling analysis is to determine the loads at which the system stiffness matrix is singular via a linear perturbation process.  In this study, we use the Lanczos algorithm to determine the eigenvalues of the system. In Abaqus, the buckling mode computation is undertaken by using the *BUCKLE input file command. Modes can then be added as imperfections to a mesh using the *IMPERFECTION input file command or by moving each node manually.

General static post-buckling is perhaps the simplest method to determine the post-buckling behavior of a wrinkled sheet and has formed the basis of several numerical studies including shearing (\citealt{WP2006c}) and uniaxial stretch (\citealt{Zheng2009}) of a thin sheet under displacement control. First, an imperfection based on some number of the computed buckling modes is added to the mesh to promote the formation of wrinkles.  As will be demonstrated in the numerical results, the wrinkle morphology predicted by the post-buckling analysis is strongly affected by the modes chosen as an imperfection. Next, a geometrically nonlinear analysis is carried out whereby the  system of equations are solved using Newton-Raphson iteration (this analysis is started using the *STATIC, NLGEOM command). The prescribed load (or displacement) is applied incrementally over a user-specified "time" interval. In order to avoid numerical instabilities, artificial viscous damping can be applied via a stabilization factor. Ideally, the damping used is kept small enough to not significantly alter the solution of the problem in the stable regime. Viscous damping is added using the STABILIZE directive.

\subsection{Finite Elements: Modified Riks Method\label{sec:FE_Riks}}
The Riks method is an arc-length continuation procedure used primarily to find solutions to problems involving global instabilities such as snap-through (in force control) or snap-back (in displacement control). These are situations where the general static method typically fails to provide solutions in the unstable regime of the force-displacement space.  The modified Riks method used in Abaqus assumes that all of the applied loads on the sheet can be scaled by a single parameter introduced via an auxiliary constraint equation. This "load proportionality factor" is computed at each step (via Newton-Raphson iteration) as part of the solution. Solution progress is recorded with respect to arc-length measured along the force-displacement path rather than the "time" parameter of the general static approach. A drawback of the Riks method is that it can be difficult to obtain the solution at exactly a target load.  User specified loads are only used to set a reference direction and magnitude. Commonly, Abaqus will return a solution near that target.

Although not specifically intended for local instabilities, Riks has been used in wrinkling studies (\citealt{Nayyar2011}). However, (\citealt{WP2006c}) tried to apply it to problems such as uniaxial tension and shear of a rectangular sheet in displacement control, but were unsuccessful.

In Abaqus, the Riks option is chosen using *STATIC, RIKS with NLGEOM. Like in the general static approach, we use initial imperfections based on the buckling eigenmodes.

\subsection{Dynamic Relaxation Method \label{sec:DR_method}}
Dynamic relaxation has been used successfully to solve a wide variety of equilibrium problems in tension structures such as membranes and cables (\citealt{Silling1988}, \citeyear{Silling1989}; \citealt{Shugar1990, HaseSteig1994,Epstein2001,TaySteig2009,Rezaiee2011,Rod2011}). Here, we briefly summarize the dynamic relaxation method implemented by \cite{Taylor2014} for the simulation of equilibrium deformations in thin sheets.

Dynamic relaxation (DR) involves embedding the equilibrium problem into an artificial dynamical system with positive-mass. As there are no known theorems for the existence of energy minimizers for (\ref{eqn:eq2}), \cite{Taylor2014} argues that this gives additional motivation for the use of dynamic relaxation as it regularizes the problem. Thus, (\ref{eqn:eq3}) is replaced by surrogate equation of motion,
\begin{equation}
\label{eqn:eq7}
T_{i\alpha,\alpha}=\rho\ddot{r}_{i}+c\dot{r}_{i},
\end{equation}
where $\rho$ is a mass density parameter and $c$ is a damping parameter.  A strength of the method is that the mass and damping parameters do not need to be representative of any physical analogues of the problem of interest.  Instead, they are chosen to maximize stability and rate of convergence to equilibrium. Optimal parameters for mass and damping for a chosen time-step size can be found using the eigenvalues of the system stiffness (or tangent stiffness) matrix (\citealt{Shugar1990,TopKhan1994,Rezaiee2011}).

Equation (\ref{eqn:eq7}) is discretized in time using central differences.  In space, the system is discretized using a finite-difference scheme derived from Green's theorem, which was originally described by Silling (\citeyear{Silling1988,Silling1989}) and adapted to membrane theory in \cite{HaseSteig1994} and \cite{TaySteig2009}. This discretization does not lead to an explicitly computed stiffness matrix as is common in conventional finite difference schemes.  Rather, it results in a system of nonlinear equations. To avoid the computational cost of computing a tangent stiffness matrix to determine optimal dynamic relaxation parameters, \cite{Taylor2014} uses an efficient variant of dynamic relaxation called kinetic damping (\citealt{Shugar1990,TopKhan1994,Rezaiee2011}).  In this method, the damping parameter is set to zero. Instead of explicit viscous damping, the kinetic energy of the system is monitored.  When it reaches a peak, the velocities of each node are set to zero and the simulation restarted. This process is continued until the system is within a user-specified tolerance of equilibrium.  Thus, the number of parameters is reduced to one (i.e., the mass).  Satisfactory values of the mass can be easily determined by hand.

\section{Case Study: Annular sheet under differential tension\label{CaseStudy}}

\begin{figure}[h]
\begin{centering}
\includegraphics[scale=0.4]{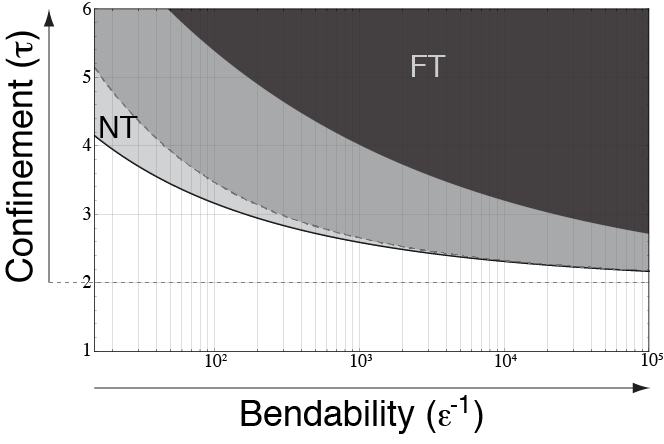}
\par\end{centering}
\caption{Phase diagram of the space spanned by bendability and confinement (reproduced from Davidovitch et al. 2011, 2012). The white region corresponds to a flat sheet. Light gray corresponds to the near-threshold (NT) regime, while the black region corresponds to the far-from-threshold (FT) regime.  The darker gray region corresponds to the transition region between NT and FT regimes.}
\label{fig:confbend}
\end{figure}

The Lam\'e set-up is the simplest (yet nontrivial) example of a radial stretching problem, whereby an annular sheet of thickness $t$ and radii $R_{in} < R_{out}$ is subjected to co-planar radial tensile loads $T_{in}$ and $T_{out}$ at its inner and outer boundaries, respectively. Intuitively, if $T_{in}$ is sufficiently larger than $T_{out}$, a region near the inner boundary is pulled inward, such that the sheet is subjected there to hoop compression, which is relieved through a wrinkle pattern. For a sufficiently thin sheet, the stress field, as well as the location, structure, and evolution of the wrinkle pattern are governed by two dimensionless groups only, $\tau$ and $\epsilon^{-1}$, which were termed, respectively, the ``confinement" and ``bendability" of the system:
\begin{equation}
\tau = \frac{T_{in}}{T_{out}} \ \ ; \ \ \epsilon^{-1} = \frac{T_{out} R_{in}^2}{B} \ ,
\end{equation}
where $B  = E t^3/[12(1-\nu^2)]$ is the bending modulus of the sheet. Fig. \ref{fig:confbend} shows a ``morphological phase diagram", spanned by the dimensionless pair $(\epsilon^{-1},\tau)$. As the load ratio $\tau$ increases, the system evolves from a flat state (white), through a NT parameter regime (light gray), to a fully developed, FT state (black), at which the wrinkle pattern can be described through perturbation theory of the singular limit of infinitely thin sheet, addressed by tension-field theory. Two features of this phase diagram are noteworthy.  First, the transition between the NT and FT regimes occurs through a narrow window in the parameters space $(\tau,\epsilon^{-1})$, that vanishes as $\epsilon \to 0$, indicating the relevance of tension field theory at confinement values arbitrarily close to threshold. Second, neither $\tau$ nor $\epsilon$ involves the sheet's stretching modulus ($Y$), implying that the strains may remain arbitrarily small (since the tensile strains scale as $T/Y$), while the wrinkle pattern evolves to its fully developed, FT state. Hence, all basic aspects of the tensional wrinkling phenomenon can be studied quantitatively within a theoretical framework that assumes Hookean material response (i.e., stress proportional to strain).

We begin with a characterization of the NT parameter regime, which is based on the exact solution of the planar state, attributed to Lam\'e. In this regime, the hoop stress in the sheet can be expressed as (\citealt{Timoshenko1970})
\begin{equation}
\sigma_{\theta\theta} = \frac{-T_{out}R_{out}^2+\tau T_{out}R_{in}^2}{R_{in}^2-R_{out}^2} + \left[\frac{-T_{out}+\tau T_{out}}{R_{in}^2-R_{out}^2}\right]\frac{R_{in}^2R_{out}^2}{r^2},
\label{eqn:hoopstress}
\end{equation}
where $r$ is radial distance measure from the inner boundary. When the confinement ratio exceeds a critical value $\tau > \tau_c \approx 2$, the hoop stress becomes compressive in the zone $R_{in}<r<L_{NT} (\tau)$, whose width increases with $\tau$. The parameter $L_{NT}$ is therefore the radial distance from the center where the hoop stress transitions from negative to positive values. Setting $\sigma_{\theta\theta}=0$ in (\ref{eqn:hoopstress}), we have (\citealt{Pineirua2011,Davidovitch2011})
\begin{equation}
L_{NT} = R_{in}\sqrt{\frac{1-\tau^{-1}}{\tau^{-1}-\frac{R_{in}^2}{R_{out}^2}}}.
\label{eqn:LNT}
\end{equation}
A linear stability analysis around the Lam\'e solution (\citealt{Coman2007b}) shows that $\tau_c - 2 \sim \epsilon^{1/4}$, and yields the scaling of the wrinkle number:
\begin{equation}
\ m_{NT} \sim \epsilon^{-3/8}    \label{NT}  \ . \\
\end{equation}
Importantly, Eq.~(\ref{NT}) shows that the number of wrinkles diverges, as the sheet becomes thinner ({\emph{i.e.}} $\epsilon \to 0$) even when the sheet is very close to threshold.

Next, we characterize the FT state, which starts with solving for the compression-free stress field in the tension-field theory (addressing the singular limit $\epsilon=0$) for a given $\tau$ (\citealt{Coman2007,Davidovitch2011}). To do this, we introduce a parameter $L_{FT}(\tau)$ that defines the annular region $R_{in}<r<L_{FT} (\tau)$ where $\sigma_{\theta\theta}=0$. To find the (non-zero) hoop stress for $r > L_{FT}$, we take (\ref{eqn:hoopstress}) with $R_{in}\rightarrow L_{FT}$ and $T_{in}-T_{out} \rightarrow \frac{T_{in}R_{in}}{L_{FT}}-T_{out}$, yielding (\citealt{Pineirua2011,Davidovitch2011})
\begin{equation}
\sigma_{\theta\theta} = \frac{-T_{out}R_{out}^2+\left(\frac{\tau R_{in}}{L_{FT}} \right)T_{out}L_{FT}^2}{L_{FT}^2-R_{out}^2} + \left[\frac{-T_{out}+\left(\frac{\tau R_{in}}{L_{FT}} \right) T_{out}}{L_{FT}^2-R_{out}^2}\right]\frac{L_{FT}^2R_{out}^2}{r^2}.
\label{eqn:hoopstressFT}
\end{equation}
Then, to find $L_{FT}$, we set the hoop stress to zero, yielding (\citealt{Pineirua2011})
\begin{equation}
L_{FT} = \frac{R_{out}}{\tau} \left[\frac{R_{out}}{R_{in}} - \sqrt{\frac{R_{out}^2}{R_{in}^2}-\tau^2}    \right].
\label{eqn:LFT}
\end{equation}
We note that $L_{FT} \neq L_{NT}$, i.e., the extent of the wrinkles in the two regimes is different. Using tools of singular perturbation theory (\citealt{Davidovitch2012,Qiu}, \textit{in preparation}), the scaling law of the wrinkle number can be derived, also exhibiting sharp difference from the NT analysis:
\begin{equation}
m_{FT} \sim \epsilon^{-1/4}.   \label{FT}
\end{equation}
In addition, the profile of the wrinkles $f(r)$ away from their tip is given by (\citealt{Davidovitch2012})
\begin{equation}
\frac{1}{4}f(r)^2m^2=\frac{R_{in}T_{out}}{Y}\tau r \text{ln}\left[\frac{\tau R_{in}}{2r} \right].
\label{eqn:profile}
\end{equation}
This expression indicates a cusp at the wrinkle's tip, which was found to be smoothed out through a boundary layer (see Figure 2b in \citealt{Davidovitch2012}) whose width vanishes very slowly, as $\epsilon^{1/6}$, and consequently clearly affects our simulations. A recent work (\citealt{Qiu}, \textit{in preparation}), employed asymptotic matching techniques to describe the boundary layer, yielding a unified analytic expression for the whole wrinkle profile, which we use for comparison with our numerical results (more details can be found in Appendix A).

Inspection of Eqs. (\ref{eqn:LNT})-(\ref{FT}) yields a profound feature of the wrinkle pattern, which is valid, in the singular limit $\epsilon \to 0$, at both NT and FT regime: The macroscale (location) feature of the wrinkle pattern is determined by the confinement parameter $\tau$, which depends only on the exerted loads; the bending modulus affects only the microscale (structure) feature of the pattern, through the bendability parameter $\epsilon^{-1}$. When the loads ratio $\tau$ is gradually increased, the transition between the NT and FT regimes (Eqs. \ref{NT} and \ref{FT}) has been studied numerically using the FvK equation ((\citealt{Davidovitch2012,Qiu} \textit{in preparation}). That study also yields the numerical prefactor in Eq.~(\ref{FT}), which we use to compare with our simulation results (see Appendix A).

\subsection{Numerical setup\label{NumericalSetup}}
In this section, we provide the details of our numerical simulations of the Lam\'e problem.  We consider the annular sheet to be made of silicone rubber with a Young's Modulus of $1 MPa$, a Poisson ratio of $0.5$, and a mass density of $1000 \frac{kg}{m^3}$.  The outer radius is $5cm$ and the inner radius is $5mm$. Sheet thickness and loading were allowed to vary to accommodate various values of bendability and confinement, but we assured that $T_{in},T_{out} \ll Y$ so that all deformations considered are properly described by a Hookean response.

\subsubsection{Finite elements}
For all finite element simulations in this work, we use a mesh of S4R elements totaling 55,816 nodes biased such that the outer and inner boundaries comprise 400 nodes circumferentially.  As a result, more elements are located near the inner boundary relative to the outer boundary. A mesh refinement study showed that this mesh density provided a good balance between precision and computational time.  In addition, a Neo-Hookean material model is used for the membrane strains.  Although the  considered deformations are linearly elastic, a Neo-Hookean model was found to be more stable numerically than Hooke's Law.  Due to the difficulty in obtaining solutions using the finite element method (discussed below), we focus these simulations primarily on the case where $\tau = 4$. Varying the sheet's bendability within this confinement facilitates exploring the NT, intermediate, and FT regimes (see Figure \ref{fig:confbend}).

All of the results shown in this study were obtained using eigenmodes as imperfections. We also attempted to use imperfections not based on the buckling modes.  In particular, we tried imperfections of the type given in Eqns (\ref{eqn:eq8}) and (\ref{eqn:eq9}) below. However, these either led to simulations that failed to converge or led to solutions with no predicted wrinkling and the wrong stress field.

We finally note that the simulations performed in this study are under force control.  This is in contrast to the majority of previous research, which has focused on displacement controlled deformations (\citealt{Zheng2009,WP2006c,Nayyar2011,Healey2013}).

\paragraph{Finite Elements: Buckling\label{sec:NS_FE_Buckle}}
For the finite element simulations, two different values of thickness were considered: $2.0 \times 10^{-6} m$ and $2.9 \times 10^{-6} m$ to achieve the desired bendabilities while remaining in the linear elastic regime.  For both values of thickness, we computed the first $100$ eigenmodes using Abaqus' built-in Lanczos solver. The differences in the modes obtained in either case are negligibly small.

In all of our analyses, buckling modes come in pairs (e.g., mode 1 and 2, 3 and 4, 5 and 6, etc.) with each member of the pair sharing a nearly identical critical load and having a shape that differs only by a small rigid rotation. The general trend is for the number of wrinkles to increase with increasing critical load; however, modes 1 and 2 have three wrinkles, while modes 3 and 4 have two.

\begin{figure}[h]
\begin{centering}
\includegraphics[scale=0.5]{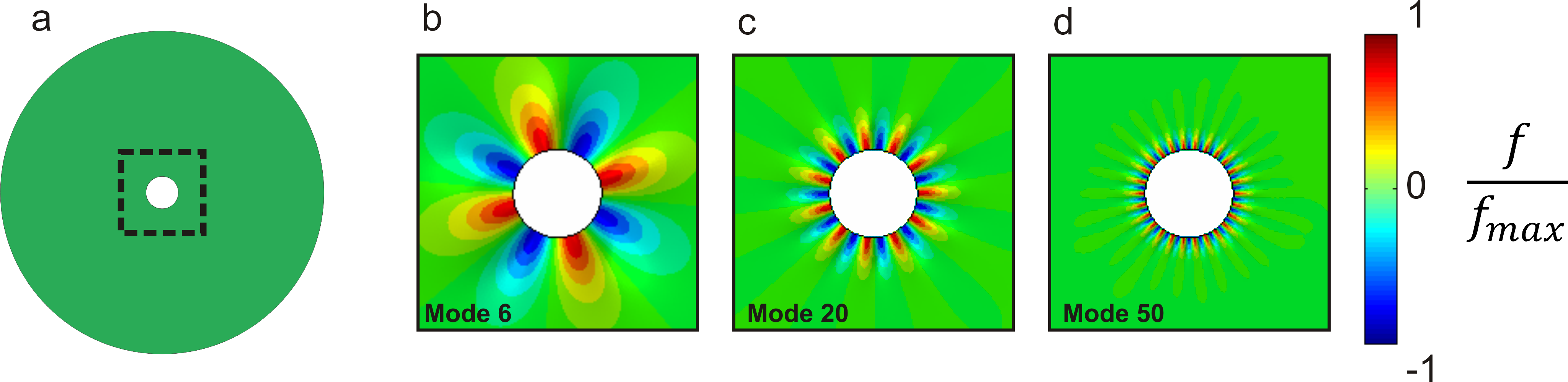}
\par\end{centering}

\caption{Schematic of buckling modes used as imperfections in this analysis. In (a), the entire annulus is shown with dashed box indicating the regions shown in (b), (c), and (d). Eigenmodes 6, 20, and 50 are shown in (b), (c), and (d), respectively and colored based on normalized out-of-plane displacement ($f$).}
\label{fig:bucklemodes}
\end{figure}

The mode shapes used as imperfections in the post-buckling analyses are shown in Figure \ref{fig:bucklemodes}. Many other combinations of modes were studied, but modes 6, 20, and 50 were found to lead to successful simulations across the widest range of bendabilities.  The critical loads for modes 6, 20, and 50 are $T_{in} = 3.04 \times 10^{-6} N/m$, $1.27 \times 10^{-5} N/m$, and $5.95 \times 10^{-5} N/m$, respectively. We also note that the eigenvalues computed by Abaqus are not normalized.  The amplitudes for the three modes as returned by Abaqus are $1.25 \times 10^{-3} m$, $4.54 \times 10^{-4} m$, and $1.89 \times 10^{-4} m$ for modes 6, 20, and 50, respectively.

\paragraph{Finite Elements: Static\label{sec:NS_FE_Static}}

Solutions using the general static analysis were very difficult to obtain. This is perhaps not surprising given that this case study is load controlled.  The primary "dials" with which to obtain a good stable solution are the number of modes used as an imperfection, the amplitude of the imperfection, and the stabilization factor.  All were varied as part of a parameter study and it was found that the ability of Abaqus to determine a solution was highly sensitive to the imperfection. The stabilization factor was not observed to make a significant difference in the range of $1 \times 10^{-8}$ (used by \citealt{Zheng2009,WP2006c}) to $2 \times 10^{-4}$ (Abaqus default).  The most consistently good results were found using a stabilization factor of $1 \times 10^{-8}$ and an imperfection mode scaling factor of $7.5 \times 10^{-4}$ (applied to the non-normalized eigenmode amplitudes).

As we discuss below, we found that wrinkled solutions can only be obtained if the chosen mode used as an imperfection represents a configuration very close to the equilibrium wrinkle pattern.  That is, information about the solution needs to be known before using a static analysis.

\paragraph{Finite Elements: Riks\label{sec:NS_FE_Riks}}

For this particular problem, the Riks method was more successful in finding solutions to the Lam\'e problem; however, solutions were still difficult to obtain and very sensitive to input parameters. A frequently encountered problem was one where Abaqus would flip the signs of both $T_{in}$ and $T_{out}$ putting the entire annulus into radial compression.  This is understandable as Riks is a method to obtain solutions for global buckling exactly of the type generated by putting the annulus in radial compression.  Nonetheless, it is still possible for Abaqus to return good solutions provided the imperfection and Riks parameters are chosen very carefully.

For the Lam\'e problem, we had the best success using an imperfection amplitude scaling of $9.0 \times 10^{-4}$ (applied to the non-normalized eigenmode amplitudes). In addition, we used the following Riks parameters: Initial Arc Increment = $1.0$, Min. Arc Increment = $0.5$, Max. Arc Increment = $2.0$, Total Arc Length = $10.0$, and Max. Number of Increments = $100$. For simulations aimed at the NT regime, we consider a target bendability of 5000 (i.e., the bendability for which the load proportionality factor is $1.0$).  When we were aiming for simulations in the FT regime, we used a target bendability of 15000.

In contrast to the static analysis, we found that the imperfections modes used in a Riks analysis do not need to be as close to the solution pattern.  Riks can generate solutions using modes very far away from the true solutions.  However, in these cases, the resulting wrinkle pattern is often asymmetric as the procedure attempts to add the necessary wrinkles over the course of the simulation. In this study, we do not show asymmetric results as they are obviously incorrect.

\subsubsection{DR Method\label{sec:NS_DR}}
The dynamic relaxation method was able to determine solutions at any desired bendability and confinement.  The user only needs to choose the starting imperfection and the virtual mass.  A stable mass can be found very easily via trial and error--larger values tending to make the simulation more stable.  The DR technique is also relatively insensitive to starting imperfections. For the user then, the choice of mass and imperfection is done mainly with the goal of promoting as fast a convergence to equilibrium as possible. To that end, a low mass and a smooth imperfection is desirable.

In this study, we use the following out-of-plane imperfections.  The first is a smooth perturbation of the form,
\begin{equation}
\label{eqn:eq8}
r_3^{t=0} = A\sin\left(\omega\frac{\pi r_1^{t=0}}{2R_o}\right)\sin\left(\frac{\pi r_2^{t=0}}{2R_o}\right),
\end{equation}
where $A = 10^{-7} m$ is the amplitude, $\omega = 5$ is a frequency, and $R_o = 5cm$ is the outer radius of the annulus.  The second is a random imperfection of the form,
\begin{equation}
\label{eqn:eq9}
r_3^{t=0} = A(2n-1),
\end{equation}
where $A$ is the same amplitude as above and $n$ is a random number between $0$ and $1$.  This random number is generated anew for each node, thus, each point has a unique initial out-of-plane position.

For all simulations, we used a radial mesh logarithmically biased such that more nodes occur near the inner boundary.  The mesh density was varied from 40,000 (500 nodes distributed circumferentially and 80 radially) nodes for simulations in the NT regime to 200,000 (200 nodes distributed circumferentially and 100 nodes radially)for those far into the FT regime.

In addition, in all simulations the dynamic relaxation procedure was terminated after 200,000 iterations (i.e., time-steps). While the stresses converge to their equilibrium values quite quickly (on the order of 10,000 iterations), the wrinkle patterns do not. We find that the finer the distribution of wrinkles in the equilibrium configuration, the longer the DR procedure takes.

\subsection{Numerical Results\label{sec:NumRes}}
In this section, we show results comparing the considered numerical methods with the analytical predictions given in section 4. To simplify references to specific finite element results, we list the post-buckling method followed by the mode used as an imperfection in parentheses.  For example, Riks (mode 20) denotes that we are referring to results generated using the Riks method with mode 20 used as the imperfection.

\subsubsection{Hoop stress \label{sec:hoop}}
\begin{figure}[H]
\begin{centering}
\includegraphics[scale=0.4]{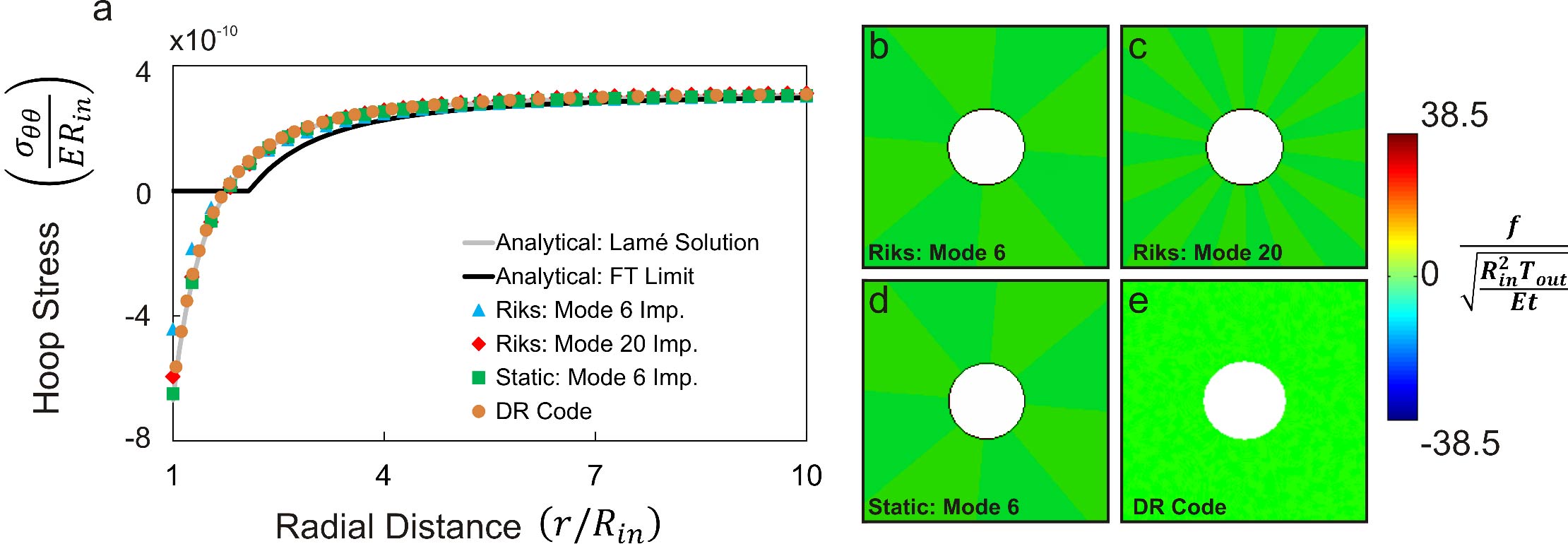}
\par\end{centering}

\caption{Comparison of hoop stress and wrinkle patterns in the case $\epsilon^{-1}=15$. In (a), the normalized hoop stress stress as a function of radial distance from the inner boundary is compared.  In this case, all numerical methods predict the classical Lam\'e solution.  In (b-e), the out-of-plane displacement ($f$) is normalized and shown in the region denoted in Figure \ref{fig:bucklemodes}a.  At this low bendability, none of the numerical methods predict wrinkling.}
\label{fig:hoop1}
\end{figure}

\begin{figure}[H]
\begin{centering}
\includegraphics[scale=0.3]{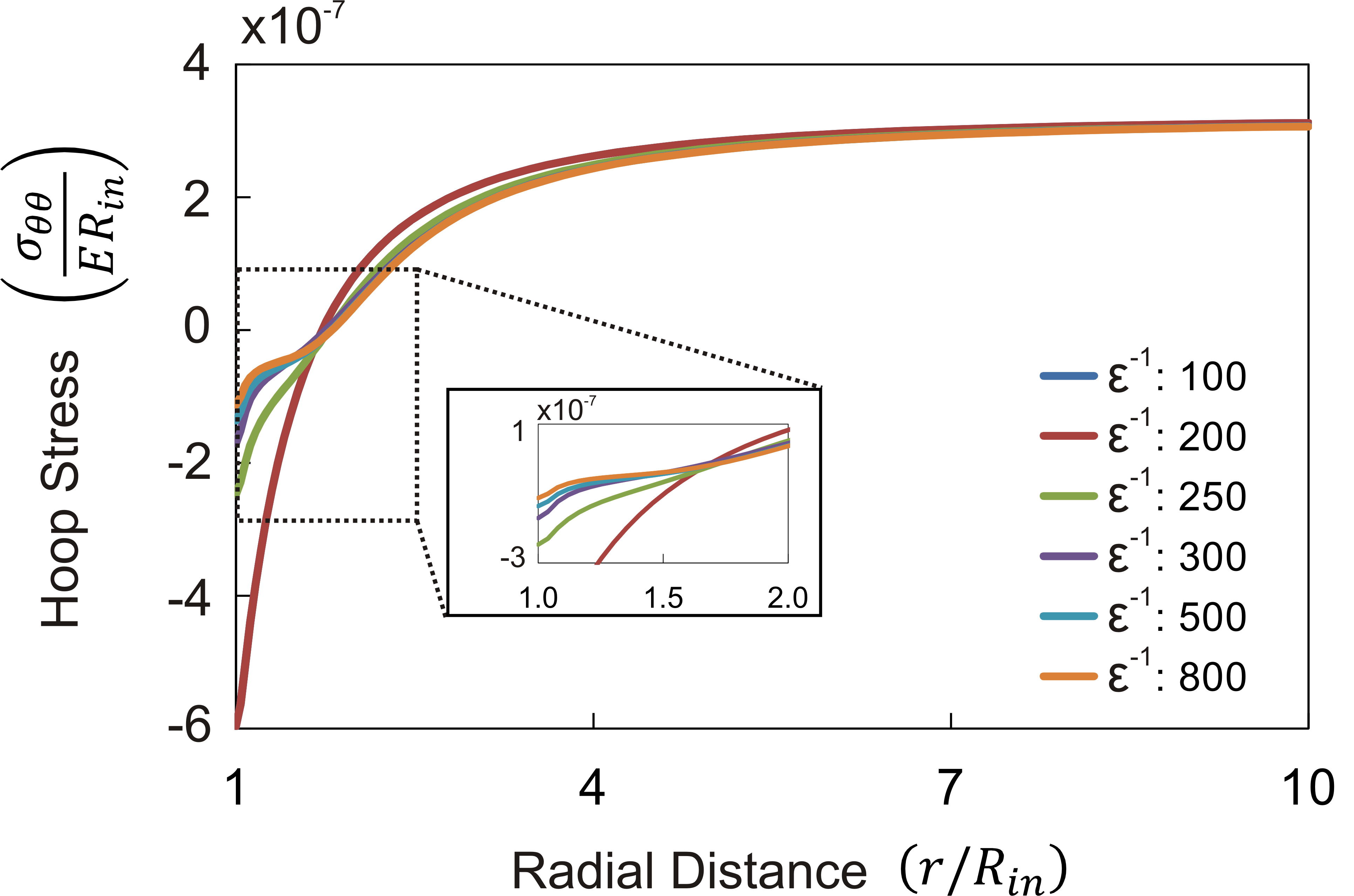}
\par\end{centering}

\caption{Comparison of hoop stresses predicted by the DR code for a range of bendabilities.  The hoop stress is shown as a function of radial distance from the inner boundary.  As bendability increases, the hoop stress transitions from the classical Lam\'e solution ($\epsilon^{-1}<250$) towards the FT limit solution.}
\label{fig:hoop4}
\end{figure}

\begin{figure}[H]
\begin{centering}
\includegraphics[scale=0.4]{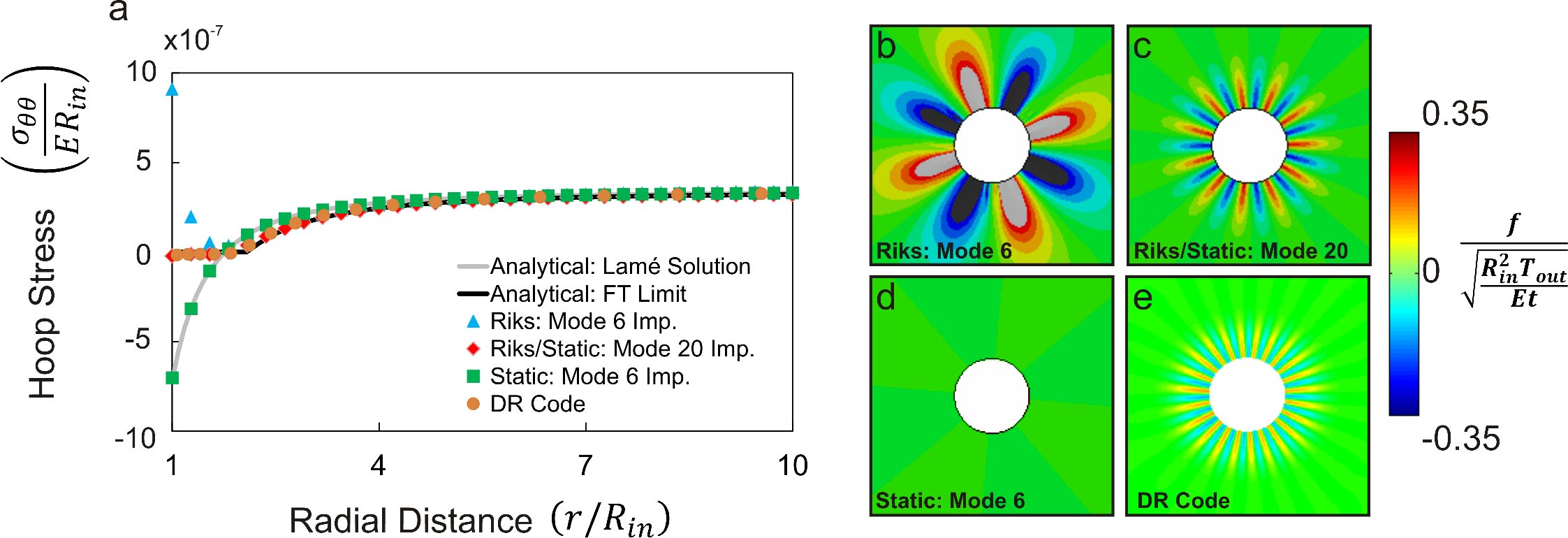}
\par\end{centering}

\caption{Comparison of hoop stress and wrinkle patterns in the case $\epsilon^{-1}=16200$. In (a), the normalized hoop stress stress as a function of radial distance from the inner boundary is compared.  The Riks, static (mode 20), and DR approaches yield solutions near the FT limit (i.e., tension-field) solution as expected. Using mode 6 with Riks causes a spurious spike in stress near the inner boundary. The general static post-buckling approach using mode 6 incorrectly predicts the Lam\'e solution.  In (b-e), the out-of-plane displacement ($f$) is normalized and shown in the region highlighted in Figure \ref{fig:bucklemodes}a. The Riks (b,c) and static (mode 20) (c) methods return the same number of wrinkles as the initial imperfection, but at a larger magnitude.  The DR method (e) predicts many more wrinkles. The general static approach (mode 6) predicts no wrinkling (d).}
\label{fig:hoop2}
\end{figure}

\begin{figure}[H]
\begin{centering}
\includegraphics[scale=0.4]{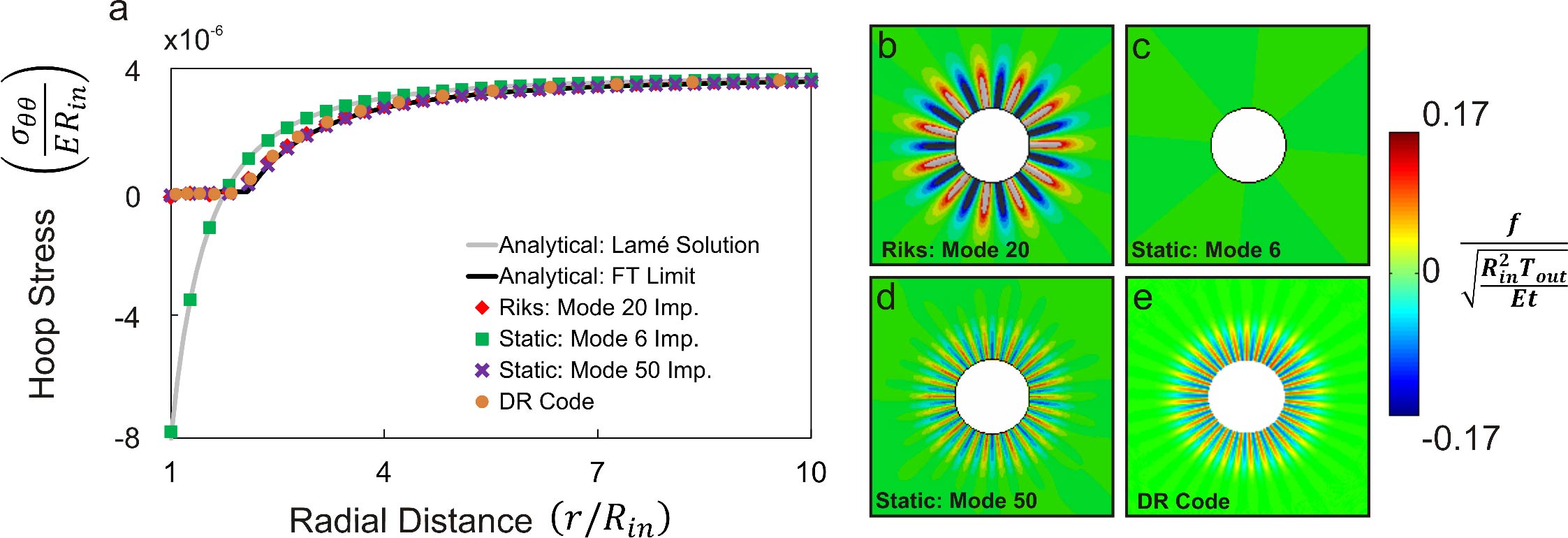}
\par\end{centering}

\caption{Comparison of hoop stress and wrinkle patterns in the case $\epsilon^{-1}=60,000$. In (a), the normalized hoop stress stress as a function of radial distance from the inner boundary is compared. The Riks, static (mode 50), and DR approaches yield solutions at the FT limit (i.e., tension-field) solution as expected. The general static post-buckling approach (mode 6) incorrectly predicts the Lam\'e solution.  In (b-e), the out-of-plane displacement is ($f$) is normalized and shown in the region highlighted in Figure \ref{fig:bucklemodes}a. When using mode 20 (b), the Riks method returns the same number of wrinkles as the initial imperfection, but at a larger magnitude. Similar behavior is observed using mode 50 in the static analysis (d). The DR method (e) predicts many more wrinkles. The general static approach (mode 6) predicts no wrinkling (c).}
\label{fig:hoop3}
\end{figure}

To begin, we look at hoop stress $\sigma_{\theta\theta}$ as a function of radius for values of bendability across the NT to FT spectrum (Figs. \ref{fig:hoop1}-\ref{fig:hoop3}) and compare with the analytical predictions given by (\ref{eqn:hoopstress}) and (\ref{eqn:hoopstressFT}).
First, we consider a sheet bendability of $\epsilon^{-1} = 15$ (thickness = $2.0 \times 10^{-6} m$, $T_{out} = 5.42 \times 10^{-7} N/m$) that is in the NT regime.  Figure \ref{fig:hoop1}a shows the hoop stress and, in this case, all numerical methods correctly predict the stress as being that of the classical Lam\'e solution. As expected, no wrinkling (Fig. \ref{fig:hoop1} b-e) is observed as the bendability is very low.

Next, we explore the intermediate region between the NT and FT regimes. In particular, we are interested in how the hoop stress evolves from the Lam\'e solution (\ref{eqn:hoopstress}) to that in the FT state (\ref{eqn:hoopstressFT}). Because of the sensitivity of the finite element approaches to initial imperfection, we focus on results from the DR method. In Figure \ref{fig:hoop4} the evolution of the hoop stress given by the DR code across a range of bendabilities.  As bendability increases, the stress near the inner boundary collapses from the NT prediction towards the FT profile. To obtain these distributions, a randomized out-of-plane imperfection of the form (\ref{eqn:eq9}) was used.  Using the smooth sinusoidal imperfection in (\ref{eqn:eq8}) led to a prediction of wrinkling being absent until bendabilities of around 1600, at odds with the analytical predictions. Our findings suggest that the DR method is somewhat sensitive to imperfections at low bendabilities, but not very sensitive at higher values. Always using a randomized imperfection is the best way to avoid these issues. At very high bendabilities, smoother imperfections can be safely used to speed up convergence.

In Figure \ref{fig:hoop2}a, we show the hoop stress for higher value of bendability that is closer to the FT limit, $\epsilon^{-1} = 16200$ (thickness = $2.0 \times 10^{-6} m$, $T_{out} = 5.42 \times 10^{-4} N/m$). Here, we expect a hoop stress distribution very close to the one given at the FT limit.  This is what we observe for the results from static (mode 20), Riks (mode 20), and from the DR method. Riks (mode 6) yields a hoop stress profile that is good away from the hole, but spikes upwards near the inner boundary. Interestingly, static (mode 6) gives exactly the Lam\'e hoop stress distribution.

Turning our attention to the wrinkle patterns (Figure \ref{fig:hoop2}b-e), we find that wrinkling is predicted by the Riks (mode 6) and (mode 20), static (mode 20), and the DR code; however, the profiles are all very different. Interestingly, the number of wrinkles and the profile computed using the finite element method is the same as the imperfection used but with a much larger amplitude. In contrast, the DR approach generates a solution very different from the imperfection used, predicting many more wrinkles at a smaller amplitude.  We demonstrate in the following section that the DR prediction for the number of wrinkles is much closer to the analytical predictions. Finally, we see that the static post-buckling approach only predicts wrinkles if the imperfection uses a mode with a number of wrinkles close to that in the true solution.

Finally, we focus on a sheet with a very high bendability of $\epsilon^{-1} = 60000$ (thickness = $2.9 \times 10^{-6} m$, $T_{out} = 6.5 \times 10^{-3} N/m$). In Figure \ref{fig:hoop3}a, we show the hoop stress for this case. Here, we expect profiles to be even closer to the predicted FT limit. Riks (mode 20), static (mode 50), and the DR method are all able to correctly predict the stress in this case. Again, static (mode 6) predicts the Lam\'e solution (with no wrinkles, Fig \ref{fig:hoop3}c). Attempts to generate a solution with static (mode 20) were unsuccessful as the simulations failed to converge for a range of imperfection magnitudes, step sizes, and stabilization factors.

Focusing on the wrinkle patterns, we observe that static (mode 50) is able to generate a wrinkled solution (Fig. \ref{fig:hoop3}d). Although, even then, the solution has the same number of wrinkles as the imperfection. Similarly, the Riks (mode 20) solution (Fig. \ref{fig:hoop3}b) has the same number of wrinkles as the imperfection. Further, we note that converged wrinkled solutions can be obtained for Riks (mode 6) and Riks (mode 20). While they both correctly predict the hoop stress field, they both predict asymmetric wrinkle patterns and we have omitted these results. As in the previous case, the DR method generates a wrinkle pattern(Fig. \ref{fig:hoop3}e) significantly different from the imperfection used and, as we demonstrate in the following sections, very close to the analytical predictions.

In summary, our numerical results indicate that both static and Riks finite element analyses are highly sensitive to initial imperfections, while the DR method can generate solutions for all bendabilities without requiring knowledge of the equilibrium configuration beforehand.

\subsubsection{Number of wrinkles \label{sec:numwrink}}
\begin{figure}[H]
\begin{centering}
\includegraphics[scale=0.5]{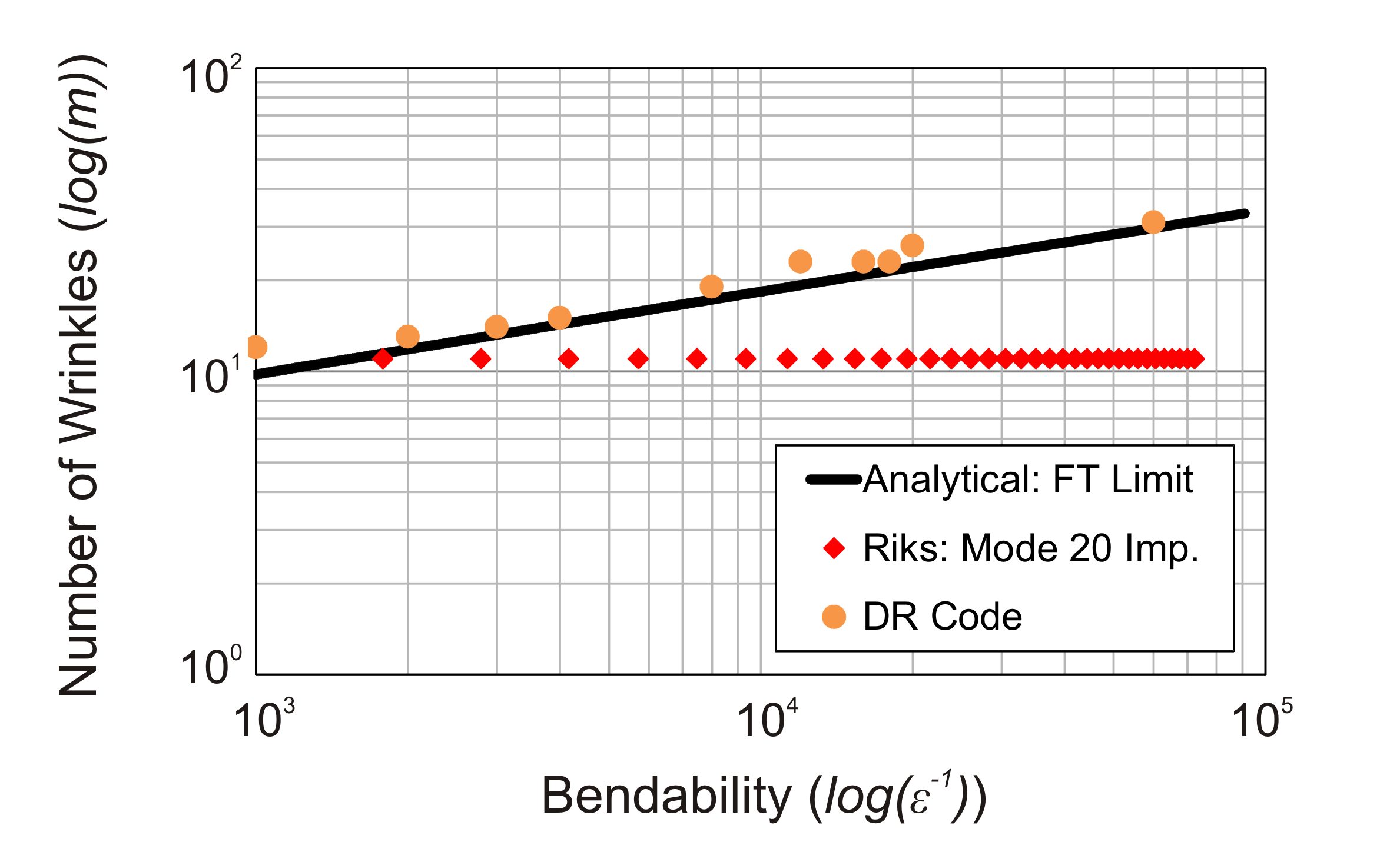}
\par\end{centering}

\caption{Comparison of number of wrinkles as a function of bendability.  The DR code yields results very close to the predicted FT limit across a range of bendabilities, while approaching the limit from above. The number of wrinkles in the Riks result is fixed at the number of wrinkles in the imperfection mode ($m_{Riks}=11$). }
\label{fig:numwrink}
\end{figure}

Next, we look at the predictions of the number of wrinkles as a function of bendability (Figure \ref{fig:numwrink}) and compare with the analytical prediction (\ref{Eq:define-c}). Here, we focus on Riks (mode 20) and the DR methods due to the difficulty in obtaining solutions with static. In a given Riks simulation, solutions at many bendabilities are provided due to the changing load proportionality factor. We see that the DR results are slightly higher than the FT limit values, which is to be expected since the analytical predictions show that the limit is approached from above (see Fig. 3 in \cite{Davidovitch2012}). It is also possible that at 200,000 iterations, the simulations are not completely converged and that at higher iterations the number of wrinkles will decrease to values closer to the FT prediction.

The Riks results under-predict the number of wrinkles and do not change as the bendability increases ($m_{Riks}=11$). This is because the mode represented has fewer wrinkles than the equilibrium solutions have. To get the proper number of wrinkles, the analyst would have to make sure to include the correct mode shape (i.e., know what it is ahead of time) as an imperfection.  Alternatively, the analyst could run many simulations each with a different initial imperfection, compute the elastic energy of the converged equilibrium state in each case, and minimize over those energies to extract the energetically-favorable mode.  This state of affairs is undesirable even in a simple problem like the one considered here.

\subsubsection{Wrinkle profiles \label{sec:prof}}
\begin{figure}[H]
\begin{centering}
\includegraphics[scale=0.35]{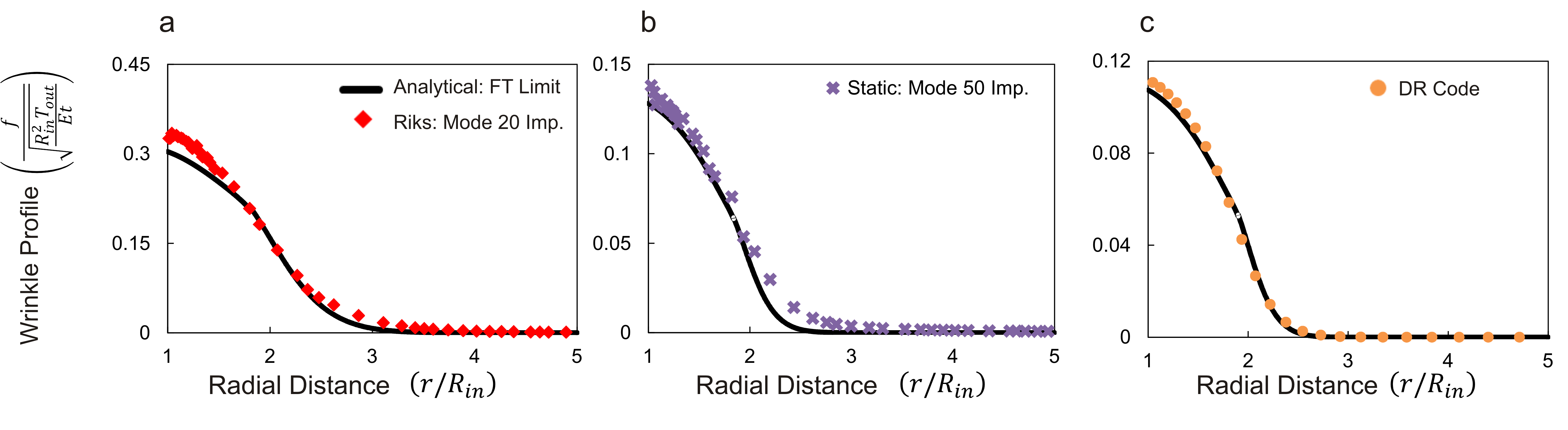}
\par\end{centering}

\caption{Comparison of averaged wrinkle profiles for ($\epsilon^{-1}=60,000$).  All profiles shown as a function of radial distance from the inner boundary. Riks (mode 20) (a), static (mode 50) (b), and the DR code (c) all give profiles close the analytical FT solution (for a particular number of wrinkles) but all predict different amplitudes}
\label{fig:prof}
\end{figure}

In Figure \ref{fig:prof} we show the out-of-plane profile of the wrinkles predicted in the case of a bendability of $\epsilon^{-1} = 60000$.  In all cases, the numerically obtained profiles have been averaged across all the wrinkles distributed circumferentially.  In addition, the analytical result (given by (\ref{Eq:appen-1})) has been scaled to account for the fact that the number of wrinkles predicted in all of the numerical results is different from the analytical prediction. The analytical prediction for the number of wrinkles in this case is 30; the DR code predicts 31, static (mode 50) predicts 26, and Riks (mode 20) predicts 11 (see Figure \ref{fig:hoop3}b,d,e). In Figure \ref{fig:prof}c, we see that the DR code predicts a wrinkle pattern nearly identical to the analytical prediction. Riks (mode 20) (Figure \ref{fig:prof}a) and static (mode 50) are also quite close to the predicted profile shape. We expect the agreement to be even better with a more refined mesh. Therefore, these results suggest that both static and Riks could generate the correct profile shape and amplitude provided the correct mode is chosen as the imperfection.

\subsubsection{Extent of wrinkles \label{sec:extent}}
\begin{figure}[H]
\begin{centering}
\includegraphics[scale=0.3]{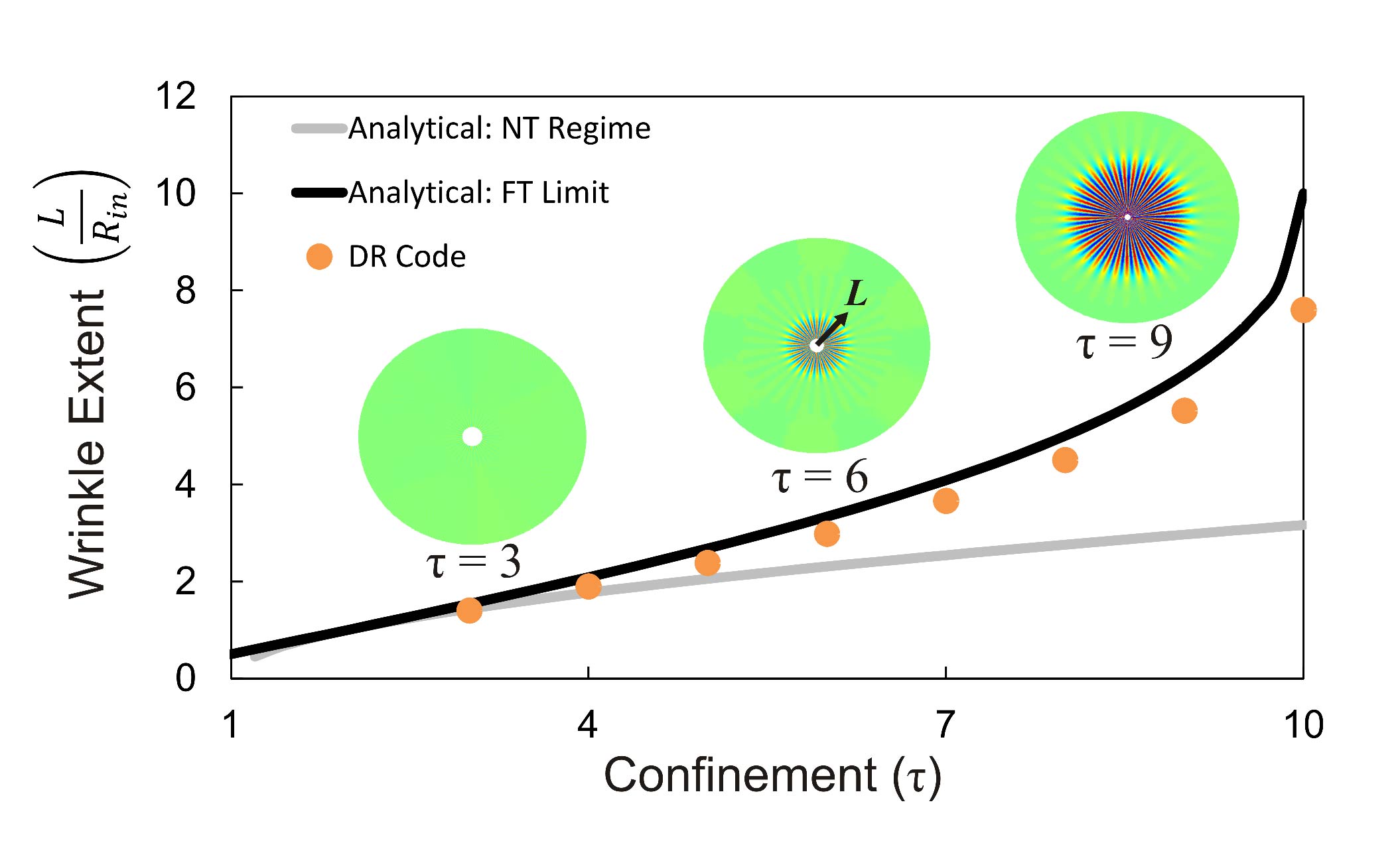}
\par\end{centering}

\caption{Comparison of wrinkle extent predicted by DR code for a range of confinements and ($\epsilon^{-1}=60,000$).  As confinement increases, the extent of wrinkles increases. At high bendabilities, the solution is expected to approach the FT limit from below and the DR results bear this out.}
\label{fig:LTau}
\end{figure}

Finally, in Figure \ref{fig:LTau}, we show the prediction of wrinkle extent, $L$, as a function of confinement for the DR method at a bendability of 60000 in comparison with the NT and FT limits predicted in (\ref{eqn:LNT}) and (\ref{eqn:LFT}) (since finite element results are very sensitive to imperfection, here we only consider the DR method). The wrinkle extent was obtained by measuring at what radial value the hoop stress went to zero. Our results clearly indicate that the DR method predicts a wrinkle extent very close to the that predicted in the FT limit over a wide range of confinement values.

\section{Conclusion\label{sec:Conclusion}}
In this work, we have compared several numerical simulation methods for solving equilibrium problems involving loaded thin sheets where wrinkling is a prominent feature. In particular, we have applied these methods to the solution of the classical Lam\'e annulus problem, which has been characterized extensively.  While the numerical methods considered are all based (at least in bending) upon Koiter's shell theory, they differ in how those shell equations are solved.  We find that the method of dynamic relaxation seems to yield the best results across the spectrum from bending dominance ('Near Threshold', NT) to membrane dominance ('Far From Threshold', FT) in capturing stress and detailed wrinkling features in comparison to analytical predictions.  The DR results appear to be dependent upon the imperfection only in the NT regime. Problems can be avoided by using a randomized imperfection in all cases.

Results based on the finite element method are much more strongly dependent on initial imperfection, when they are able to be obtained at all. The Riks method appears to be superior to the static post-buckling method since its dependence on imperfection is less severe. However, neither static nor Riks is able to adequately capture the wrinkle morphology unless the appropriate buckling mode is known \textit{a priori} and used as the imperfection. For most real applications, this requirement is much too restrictive.

In conclusion, numerical study of the Lam\'e set-up, where analytic predictions are available, reveals the power and the limitations of the finite element method in characterizing wrinkle patterns. Both the Riks and static approaches can be used to reliably determine the macro-scale properties (overall deformation, location of wrinkles, stress field).  On the other hand, the micro-scale properties (wrinkle periodicity and profile) can not be captured reliably.  Our findings suggest a dynamic approach like the DR method can more robustly predict both macro- and micro-scale properties of wrinkled sheets and demonstrate the special care that must be taken when analyzing numerical results on wrinkling phenomena.

Acknowledgment: MT and KB acknowledge the support of the Harvard Materials Research Science and Engineering Center under National Science Foundation Award DMR-0820484 and of startup funds from the School of Engineering and Applied Sciences, Harvard. ZQ and BD acknowledge the support of NSF CAREER award DMR-11-51780.

\appendix
\section{Analytic predictions for $m$ and  $f(r)$}

\newcommand \simg{\stackrel{>}{\sim}}
\newcommand \siml{\stackrel{<}{\sim}}
\newcommand \bbu{{$\bullet$}}
\newcommand \tDel{{{\tilde{\Delta}}}}
\newcommand \tx{{{\tilde{x}}}}
\newcommand \ty{{{\tilde{y}}}}
\newcommand \tq{{{\tilde{q}}}}
\newcommand \tL{{{\tilde{L}}}}
\newcommand \q{{{\theta}}}
\newcommand \tr{{{\tilde{r}}}}
\newcommand \tu{{{\tilde{u}}}}
\newcommand \tzeta{{{\tilde{\zeta}}}}
\newcommand \tT{{{\tilde{T}}}}
\newcommand \tE{{{\tilde{E}}}}
\newcommand \tsigma{{{\tilde{\sigma}}}}
\newcommand \tur{{{\tilde{u}_r}}}
\newcommand \ba{{{\bar{a}}}}
\newcommand \br{{{\bar{r}}}}
\newcommand \brf{{{\bar{f}}}}
\newcommand \beq{\begin{equation}}
\newcommand \eeq{\end{equation}}


%
In this appendix, we provide the analytic predictions for the number of wrinkles, $m$, and for the wrinkle profile, $f(r)$, plotted in black curves in Figures 8 and 9, respectively. These predictions are obtained by solving, using an asymptotic matching method, the set of nonlinear equations (4a-4d in \cite{Davidovitch2012}). Assuming that the wrinkle shape corresponds to an out-of-plane displacement $f(r)\cos(m\theta)$, it was shown in   \cite{Davidovitch2012} that the FvK equations reduce in the large bendability limit ($\epsilon \ll 1$) to this set of nonlinear ordinary differential equations (ODE's), which couple $f(r)$ to the in-plane radial displacement $u_r(r)$. Importantly, these equations resolve the unphysical singularity of the derivative $f'(r)$ at the wrinkle's tip ($r \to R_{in} \tau/2 $), which is evident in Eq.~(21), yielding there a boundary layer, whose width vanishes slowly as $\epsilon \to 0$. This nonlinear set of ODE's was solved numerically in \cite{Davidovitch2012} and the energetically-favorable wrinkle number $m$ was obtained by computing the elastic energy and minimizing it over all values of wrinkle number. In \cite{Qiu} (\textit{in preparation}), it was found that this set of equations can be solved analytically by employing a standard method of singular perturbation theory, whereby the wrinkle profile $f(r)$ is described by matching the profile $f(r)$ in Eq.~(21) to an appropriate ``inner zone" around the wrinkle's tip, which describes the boundary layer. The matching procedure is rather cumbersome, and will be described in detail elsewhere (\citealt{Qiu}, \textit{in preparation}). Here, we only provide the analytic expressions for the profile $f(r)$ and the wrinkle number $m$, that were obtained by this method, which we use for plotting the theoretical predictions in Figs.~8 and 9.

In order to simplify the complicated following expressions, we define the dimensionless $\br$ and $\brf(\br)$ as follows:
\begin{align} \br =\frac{r}{R_{in}} \ \ ,& &\brf=\frac{f}{\sqrt{R_{in}^2T_{out}/Et}} \ , \end{align}
and define the parameter $c$ through: 
%
%
\begin{equation}
m \equiv c \cdot \epsilon^{-1/4}
\label{Eq:define-c}
\end{equation}  
We find that the profile $\brf(\br)$ 
is described by the following expression: 

\begin{equation}
\begin{array}{cc}
1<\br < \frac{\tau}{2} + x \cdot \epsilon^{1/6}  \ : \ & \ \ \brf(\br) \approx 
2 \sqrt{\frac{\br \tau  \sqrt{\epsilon } \log \left(\frac{\tau }{2 \br}\right)}{c^2}} + a\cdot \text{Ai}\left(-2^{4/3}  \left(\br-\frac{\tau }{2}\right) (\frac{c^2}{\sqrt{\epsilon } \tau ^3 \left(1-\nu^2\right)})^{1/3}\right)
\\ \\
\br > \frac{\tau}{2} + x \cdot \epsilon^{1/6}  \  : \ & \brf(\br) \approx 
b \cdot\text{Ai}\left(2 \left(\br-\frac{\tau }{2}\right) (\frac{c^2}{\sqrt{\epsilon } \tau ^3})^{1/3}\right)
\label{Eq:appen-1}
\end{array}
\end{equation} 
where $\tau,\epsilon, \nu$ are, respectivley, the confinement, inverse bendability, and Poisson ratio (see main text), $\text{Ai}(\cdot)$ is the Airy function, and the parameters $x,a,b$ are given by: 
\begin{eqnarray}
x=\frac{(0.0232769 \nu ^2-0.000918511 \nu -0.250298)\tau}{c^{2/3}} \\ \nonumber \\
a=\frac{\left(\nu ^2-1\right) \epsilon^{1/3}  \left(32 c^2 \tau  x^3+\tau ^4\right)}{16 c^3 \sqrt{\tau } \left(\nu ^2-3\right) \left(-x\right)^{5/2} \text{Ai}\left(-2^{4/3} (-\frac{c^2}{\tau ^3 \left(\nu ^2-1\right)})^{1/3} x\right)} \\ \nonumber \\
b=\frac{\sqrt{\tau } \epsilon^{1/3} \left(64 c^2 x^3+\tau ^3 \left(\nu ^2-1\right)\right)}{16 c^3 \left(\nu ^2-3\right) \left(-x\right)^{5/2} \text{Ai}\left(2 (\frac{c^2}{\tau ^3})^{1/3} x\right)}
\end{eqnarray}
The expression (\ref{Eq:appen-1}) is plotted in the black curves in Fig.~9.  
%

Next, the elastic energy is evaluated (see \cite{Davidovitch2012}), yielding an algebraic equation for the wrinkle number, $m(\epsilon,\tau)$, that minimizes the energy for given $\epsilon$ and $\tau$.
 %
This algebraic equation is rather complicated and cannot be solved in a closed analytic form. We give the implicit expressions for the parameter $c(\epsilon,\tau)$ (see (\ref{Eq:define-c})) for the two values $\nu=1/3$ and $\nu=1/2$. 
For $\nu=1/3$:
\begin{align}
{c}^4 (2-1.69315 \tau +\tau  \log (\tau ))+0.0416667 \tau ^2 (\log (\epsilon )-4 \log ({c}))-0.861147 \tau ^2\nonumber \\
-0.125 \tau ^2 \log ^2(\tau )+0.673287 \tau ^2 \log (\tau )-0.25 \tau ^2 \log (\log (\tau )-0.693147)=0
\end{align}
For $\nu=1/4$
\begin{align}
{c}^4 (2-1.69315 \tau +\tau  \log (\tau ))+0.0416667 \tau ^2 (\log (\epsilon )-4 \log ({c}))-0.866029 \tau ^2\nonumber \\
-0.125 \tau ^2 \log ^2(\tau )+0.673287 \tau ^2 \log (\tau )-0.25 \tau ^2 \log (\log (\tau )-0.693147)=0
\end{align}

\bibliographystyle{elsarticle-harv}
\nocite{*}
\bibliography{Lame_References}

\end{document}